\documentclass[a4paper,12pt]{amsart}
\usepackage{amsmath,amsfonts,amssymb,cite}
\usepackage[english]{babel}

\let\n\nonumber

\newcommand{\sn}{\,{\rm sn}}
\let\bsy\boldsymbol
\let\phi\varphi
\let\p\partial

\let\le\leqslant

\let\t\tilde
\let\ds\displaystyle

\newtheorem{theorem}{Theorem}
\newcounter{rem}
\newcommand{\rem}{\addtocounter{rem}{1}\noindent\textbf{Remark \therem.} }
\numberwithin{equation}{section}
\begin{document}
\baselineskip 6.5mm
\thispagestyle{empty}

\vspace{1cm} \centerline{\large  \textbf{Vector hyperbolic equations on the sphere }}
\vskip0.2cm 
\centerline{\large  \textbf{  possessing  integrable third order symmetries} }

\vskip1cm \hfill
\begin{minipage}{13.5cm}
\baselineskip=15pt
{\bf A Meshkov ${}^{1}$
 and
   V Sokolov ${}^{2}$} \\ [2ex]
{\footnotesize
${}^1$ Orel State University, Russia
 \\
${}^{2}$ Landau Institute for Theoretical Physics, Moscow, Russia }\\

\end{minipage}

\title{}

\thanks{\rm This work was supported in part by the Russian Foundation for Basic Research (Grant No 11-01-00341-a), and by 
the Program for Supporting Leading Scientific Schools (Grant No NSh-6501.2010.2).}

\date{}
\begin{abstract}
The complete lists of vector hyperbolic equations on the sphere that have integrable third order vector isotropic and anisotropic symmetries are presented. Several new integrable hyperbolic vector 
models are found. By their integrability we mean the existence of 
vector B\"acklund transformations depending on a parameter. For all new equations such transformations are constructed. 
\end{abstract}
\maketitle
\section{Introduction}

The symmetry approach to classification of integrable PDEs  (see, for instance, \cite{sokshab, mikshyam, mss, miksok} ) is very efficient for evolution PDEs
with one spatial variable.
It is based on the existence of local higher infinitesimal symmetries and/or conservation laws. 

For hyperbolic equations
\begin{equation}\label{hyper}
u_{xy}= \Psi(u, u_{x}, u_{y})
\end{equation}
the symmetry approach assumes the existence of both $x$-symmetries of the form
\begin{equation}\label{symm1}
u_{t}=A(u, u_{x}, u_{xx}, \dots, ),
\end{equation}
and  $y$-symmetries of the form
\begin{equation}\label{symm2}
u_{\tau}=B(u, u_{y}, u_{yy}, \dots, ).
\end{equation}
For example, the famous integrable sin-Gordon equation 
$$
u_{xy}=\sin{u}
$$
admits the symmetries
$$
u_{t}= u_{xxx}+\frac{1}{2} u_{x}^{3},\qquad u_{\tau}= u_{yyy}+\frac{1}{2}
u_{y}^{3}.
$$
Since the sin-Gordon equation is invariant with respect to the interchanging  $x \leftrightarrow y$ the symmetries are defined by the same (up to  $x \leftrightarrow y$,  $t \leftrightarrow \tau $) 
equation. This evolution  equation has infinitely many  conservation laws  and therefore consists in the list of integrable equations of the form 
\begin{equation}\label{kdv}
u_{t}= u_{xxx}+\Phi(u, u_{x}, u_{xx}),
\end{equation}
obtained in \cite{ssvin,svsok1} (see \cite{ufa} for details and proofs).

In the general classification problem, all the functions $\Psi, A,B$ in (\ref{hyper}) -- (\ref{symm2}) are unknown. They have to be found from compatibility conditions for
(\ref{hyper}), (\ref{symm1}) and (\ref{hyper}), (\ref{symm2}). Such complete classification of integrable equations (\ref{hyper}) turned out to be an extremely difficult problem. 
It is still unsolved although some partial results were obtained in \cite{zibshab1,zibshab2,zibsok}. 

However if we fix somehow functions $A$ and $B,$ then it is not difficult to 
verify whether or not the corresponding function $\Psi$ exists. In particular, to find integrable hyperbolic equations of the sin-Gordon type,
one can assume \cite{MS} that both symmetries (\ref{symm1}) and (\ref{symm2}) are {\it integrable} equations of the form (\ref{kdv}).

Our goal is to apply this idea for finding interesting integrable vector hyperbolic equations of the form
\begin{equation}\label{h1}
{\bsy u}_{xy}= h_0 {\bsy u}+h_1  {\bsy u}_x+h_2  {\bsy u}_y
\end{equation} 
on the sphere. Here  ${\bsy u}$   is an $N$-dimensional (or even infinite-dimensional) vector and $h_{i}$ are some  scalar-valued  functions 
depending on two different scalar products  $(\cdot\, ,\cdot)$ and $\langle\cdot\, ,\cdot\rangle$ between vectors ${\bsy u}, {\bsy u}_x$ and ${\bsy u}_y$.  The 
sphere is defined by the equation  ${\bsy u}^2=1$.  Here and in the sequel for any vector ${\bsy a}$ we denote 
$\bsy a^2=(\bsy a,\bsy a)$.

For equations on the sphere the constraint  ${\bsy u}^2=1$  implies relations  $(\bsy u,\bsy u_{x})=(\bsy u,\bsy u_{y})=0$ and $(\bsy u,\bsy u_{xy})=-(\bsy u_x,\bsy u_{y})$. From these identities 
it follows that  equation (\ref{h1})  has the form 
\begin{equation}\label{sph_eq}
{\bsy u}_{xy}=h_1  {\bsy u}_x+h_2  {\bsy u}_y-(\bsy u_x,\bsy u_{y}) {\bsy u}. 
\end{equation}
Equation (\ref{sph_eq}) is called {\it isotropic} if the coefficients involve the scalar product  $(\cdot\, ,\cdot)$ only. In this case the $h_i$  are functions in three  scalar variables
$(\bsy u_x,\bsy u_{x}),(\bsy u_x,\bsy u_{y}), (\bsy u_y,\bsy u_{y})$. It is clear that isotropic models are invariant with respect to 
the group $SO(N)$. 

If the coefficients essentially depend on the both scalar products then the equation is called {\it anisotropic}.  In the anisotropic case the  $h_i$ are functions of nine scalar variables: 
$$
(\bsy u_x,\bsy u_{x}),\,\,\,  (\bsy u_x,\bsy u_{y}),\,\,\,  (\bsy u_y,\bsy u_{y}), \qquad \langle \bsy u,\bsy u \rangle, \,\,\,  \langle\bsy u,\bsy u_x\rangle,\,\,\,  \langle \bsy u_x,\bsy u_x \rangle, \,\,\, 
 \langle \bsy u,\bsy u_y \rangle, \,\,\,  \langle \bsy u_y,\bsy u_y \rangle, \,\,\,  \langle\bsy u_x,\bsy u_y\rangle.
$$ 
All these scalar products will be considered as 
 {\it independent variables} (cf. \cite{meshsok}).  As a result all models found in the paper are integrable for any dimension $N$.

Integrable isotropic and anisotropic vector evolution equations 
on the sphere have been studied in \cite{meshsok, BalMesh}. This paper is based on results obtained there.

The symmetry approach to integrability of vector evolution equations 
$$
 {\bsy u}_{t}=
f_m \,  {\bsy u}_k+f_{m-1}\,
 {\bsy u}_{k-1}+\cdots+f_1\, {\bsy u}_x+f_0\, {\bsy u}, \qquad   {\bsy u}_{i}=\frac{\p^i  {\bsy u}}{\p x^i},
 $$
whose coefficients $g_i$ depend on scalar products between vectors $ {\bsy u},...,{\bsy u}_{k}$, has been developed in \cite{meshsok}. By integrability of such equations we mean the
existence of infinite series of vector symmetries 
$$
 {\bsy u}_{\tau_{k}}=g_k \,  {\bsy u}_k+g_{k-1}\,
 {\bsy u}_{k-1}+\cdots+g_1\, {\bsy u}_x+g_0\, {\bsy u}
$$
and/or infinite series of conservation laws
$$
\frac{\p \rho_i}{\p t}=\frac{\p \sigma_i}{\p x},
$$
where $\rho_i, \sigma_i$ are functions of the scalar products.

In particular,  in \cite{meshsok} necessary  integrability
conditions for equations of the form 
\begin{equation}\label{evec}
  {\bsy u}_{t} =  {\bsy u}_{xxx}+f_2  {\bsy u}_{xx}+f_1  {\bsy u}_x+f_0  {\bsy u}
\end{equation}
have been found. 
It is remarkable that these conditions look very similar to the integrability conditions (see \cite{ibshab,sokshab}) 
for scalar equations  (\ref{kdv}).   
Using these integrability conditions  we obtained a complete list of isotropic  equations of form  (\ref{evec}) on the sphere that have infinite series of conservation laws. Moreover, some examples of integrable anisotropic 
 equations were presented there. One of these examples is the following equation \cite{golsok}:
\begin{equation}
{\bsy u}_t=\Big({\bsy u}_{xx}+\frac{3}{2} ( {\bsy u}_x, \ {\bsy u}_x ) {\bsy u} \Big)_x +\frac{3}{2} \langle {\bsy u}, {\bsy u}\rangle {\bsy u}_x, \qquad {\bsy u}^2=1.
\label{LL}
\end{equation}
Here $\langle {\bsy a}\, ,{\bsy b}\rangle=({\bsy a},\, R{\bsy b})$, where $R$ is an arbitrary constant symmetric  matrix. One can
assume that $R=\text{diag}(r_{1},\dots,r_{N}).$  Equation (\ref{LL}) has a Lax pair whose spectral parameter
lies on an algebraic curve of genus $1+(N-3) 2^{N-2}$. If $N=3$, then (\ref{LL}) is a symmetry for the famous Landau-Lifshitz equation. 

All integrable anisotropic equations of the form (\ref{evec}) on the sphere have been found in \cite{BalMesh}. Equations  without any constraints on the 
length of the vector ${\bsy u}$ were investigated in \cite{meshsok1, Bal,Bal2}.

Complete list of both isotropic and anisotropic equations  on the sphere that have infinitely many conservation laws is presented in Appendix 1. 
In Section 2 we find all hyperbolic equations (\ref{h1}) that have symmetries equivalent to evolution equations from  Appendix 1. In Section 3 
we show that these hyperbolic equations are vector generalizations of known scalar integrable equation \cite{zibsok,bz} 
\begin{equation}\label{sn}
u_{xy}= \sn(u) \sqrt{\vphantom{u_y^2}u_x^2+1}\sqrt{u_y^2+1}
\end{equation}
and investigate their degenerations.

To justify the integrability of equations from Section 2 we present in Section 4 auto-B\"acklund transformations with the spectral parameter for them. 
It would be interesting to find Lax representations and to develop a Hamiltonian formalism for these equations.

\vskip.3cm
\noindent
{\bf Acknowledgments.}
The authors are grateful to V. Adler, R. Yamilov, and M. Semenov-tian Shansky  for useful discussions. 
The research was partially supported by the RFBR grant 11-01-00341-a.  

\section{Classification statement}

Two equations (\ref{h1}) on the sphere ${\bsy u}^2=1$ are said to be {\it equivalent} if they are related by a composition of the following point transformations:
 
1. The scaling transformations of the form:
$$
x'=\alpha x,\qquad y'=\beta  y;
$$

2. The interchanging $y\leftrightarrow  x$;

3. The  transformations of the second scalar product  
\begin{equation}\label{metr}
\langle   u,  v\rangle\to \alpha \langle  u,  v \rangle+\beta  ( u, v), 
\end{equation}
where $\alpha $  and $\beta $ are arbitrary constants. 
In general the constants in the equivalence transformations assumed to  be complex.

Using the equivalence transformation, we can bring any integrable $x$-symmetry of the form (\ref{evec}) to one of equations from Appendix 1. 
It is important to notice that we cannot reduce both $x$ and $y$ symmetries to equations of Appendix 1 simultaneously. So we may assume only that $x$-symmetry belongs 
to the lists of Appendix 1 and $y$-symmetry is {\it equivalent} to one of equations from the lists.

\begin{theorem}
If equation (\ref{sph_eq}) possesses isotropic or anisotropic symmetries of the form 
\begin{align}\label{sym1}
{\bsy u}_t &=  {\bsy u}_{xxx}+f_2  {\bsy u}_{xx}+f_1  {\bsy u}_x+f_0  {\bsy u}\\
\intertext{and}
\label{sym2}
{\bsy u}_{\tau}& =  {\bsy u}_{yyy}+g_2  {\bsy u}_{yy}+g_1  {\bsy u}_y+g_0  {\bsy u}
\end{align}
that have infinitely many conservation laws, then it  is equivalent to one of the following equations: 
\begin{align}
\bsy u_{xy}&=\sqrt{\vphantom{\rule{0pt}{.86em}}1-\bsy u_y^2}\sqrt{\vphantom{\rule{0pt}{.79em}}1+c\,\bsy u_x^{-2}}\ \bsy u_x-(\bsy u_x,\bsy u_y)\bsy u,  \label{a01}\\[2mm]
\bsy u_{xy}&=\frac{\langle\bsy u,\bsy u_y\rangle}{\langle\bsy u,\bsy u\rangle}\bsy u_x+
\sqrt{\vphantom{\bsy u_y^2}\langle\bsy u,\bsy u \rangle-\bsy u_x^{2}}\sqrt{\vphantom{\bsy u_x^2}1+c\,\langle\bsy u,\bsy u\rangle\,\xi^{-1}}\ \bsy u_y-(\bsy u_x,\bsy u_y)\bsy u,    \label{a12}\\[2mm]
&\xi=\langle\bsy u,\bsy u\rangle \langle\bsy u_y,\bsy u_y\rangle-\langle\bsy u,\bsy u_y\rangle^2, \n 
\end{align}
\begin{align}
\bsy u_{xy}&=\frac{\bsy u_y}{\langle\bsy u,\bsy u\rangle}\Big(\langle\bsy u,\bsy u_x\rangle+\sqrt{\xi}\,\sqrt{1+c\,\langle\bsy u,\bsy u\rangle\bsy u_y^{-2}}\,
\Big)-(\bsy u_x,\bsy u_y)\bsy u,                                                                                                                                                                                    \label{a15}\\[2mm]
&\xi=\langle\bsy u,\bsy u_x \rangle^2+\langle\bsy u,\bsy u\rangle\big(1 -\langle\bsy u_x,\bsy u_x \rangle\big), \n 
\end{align}
\begin{align}
\bsy u_{xy}&=\frac{\langle\bsy u,\bsy u_y  \rangle}{\eta}\bsy u_x
+\frac{\bsy u_y}{\langle\bsy u,\bsy u\rangle}\Big( \langle\bsy u,\bsy u_x\rangle+ \sqrt{\vphantom{\rule{0pt}{.89em}}\psi}\,
\sqrt{\vphantom{\rule{0pt}{.79em}}1+a\,\eta\,\langle\bsy u,\bsy u\rangle\,\mu^{-1}}\Big)                                                              \label{a16}
-(\bsy u_x,\bsy u_y)\bsy u,              \\[2mm]
&\psi=\langle\bsy u,\bsy u_x \rangle^2+\langle\bsy u,\bsy u\rangle(\eta- \langle\bsy u_x,\bsy u_x \rangle),\ \ \eta=\langle\bsy u,\bsy u\rangle+b,\n\\
&\mu=\eta\,(b\,\bsy u_y^2+\langle\bsy u_y,\bsy u_y\rangle)-\langle\bsy u,\bsy u_y\rangle^2, \n
\end{align}
\end{theorem}

{\bf Proof.}
Let us denote the right hand sides of (\ref{sph_eq}) and  (\ref{sym1}) by $\bsy H$ and ${\bsy K}$. It follows from the constraint $\bsy u^2=1$ and its consequences 
$(\bsy u,\bsy u_t)=0,\ (\bsy u,\bsy u_{xy})=-(\bsy u_x,\bsy u_{y})$ that
$$
\bsy H=h_1\bsy u_x+h_2\bsy u_y-(\bsy u_x,\bsy u_y)\,\bsy u \ \ \text{ and } \ \
\bsy K=\bsy u_{xxx}+f_{2}(\bsy u_{xx}+\bsy u_x^2\,\bsy u)+f_{1}\bsy u_{x}+3\,(\bsy u_x,\bsy u_{xx})\,\bsy u.
$$

The compatibility condition for  equations $\bsy u_{xy}=\bsy H$ and $\bsy u_t=\bsy K$ has the form
\begin{equation}\label{opreq1}
\frac{d^{2}}{d x dy}\bsy K= \frac{d}{dt} \bsy H ,
\end{equation}
where $d/dx,\ d/dy$ and $d/dt$ are the total derivatives.  All mixed derivatives $\bsy u_{xy}$, $\bsy u_{xxy}$, $\bsy u_{xyy}, \dots$ 
have to be eliminated from (\ref{opreq1}) in virtue of (\ref{sph_eq}) whereas to eliminate the derivatives $\bsy u_t, \bsy u_{xt}$, $\bsy u_{yt}, \dots$ we should use (\ref{sym1}). 
The remaining vector variables $\bsy u, \bsy u_x, \bsy u_{xx}, \dots , \bsy u, \bsy u_y, \bsy u_{yy}, \dots$ and their scalar products 
\begin{equation}\label{dynvard}
\begin{aligned}
&u_{x[i,k]}=(\bsy u_{x_i},\bsy u_{x_k}),\ u_{y[i,k]}=(\bsy u_{y_i},\bsy u_{y_k}),\ 0<i\le k;\ u_{xy[i,k]}=(\bsy u_{x_i},\bsy u_{y_k}),\ i>0,\ k>0, \\
&\t u_{x[i,k]}=\langle\bsy u_{x_i},\bsy u_{x_k}\rangle,\ \t u_{y[i,k]}=\langle\bsy u_{y_i},\bsy u_{y_k}\rangle,\ 0\le i\le k;\ \t u_{xy[i,k]}=\langle\bsy u_{x_i},\bsy u_{y_k}\rangle,\ i>0,\ k>0
\end{aligned}
\end{equation}
have to be regarded as independent variables.

Similarly, the compatibility condition for equations $\bsy u_{xy}=\bsy H$ and $\bsy u_{\tau}=\bsy S$ has the form
\begin{equation}\label{opreq2}
\frac{d^{2}}{d x dy}\bsy S= \frac{d}{d \tau} \bsy H.
\end{equation}

If the right hand side $\bsy K$ of the $x$-symmetry is fixed we may split (\ref{opreq1}) with respect to all vector and scalar independent variables containing third and second derivatives to get an overdetermined 
system of PDEs for $h_1$ and $h_2$.
We are going to take one by one all equations from Appendix 1 for $x$-symmetries (\ref{sym1}). 

Let us take for $\bsy K$ the right hand side of the equation 
\begin{equation}\label{A01}
\bsy u_t= \bsy u_{xxx}-3\,\frac{(\bsy u_x,\bsy u_{xx})}{\bsy u_x^2}\,\bsy u_{xx}+\frac{3}{2}\,\biggl(\frac{\bsy u_{xx}^2}{\bsy u_x^2}+ 
\frac{(\bsy u_x,\bsy u_{xx})^2}{\bsy u_x^4\,(1+a\,\bsy u_x^2)}\biggr)\,\bsy u_x, 
\end{equation}
labeled (\ref{A1}) in Appendix 1. It is easy to verify that the coefficient at the highest vector derivative 
 $\bsy u_{xxx}$ in (\ref{opreq1}) has the form 
\begin{equation}\label{U3}
\begin{aligned}
&\frac {3}{u_{x[1,1]}^{2}} \left[u_{x[1,1]}u_{xy[2,1]}\left(u_{xy[1,1]}\frac{\p h_2}{\p u_{xy[1,1]}}+h_2\right)+2\,u_{xy[1,1]}u_{x[1,2]}\left(u_{x[1,1]} \frac{\p h_2}{\p u_{x[1,1]}}-h_2\right)\right]\\
&+\frac {3\,u_{xy[1,1]}}{u_{x[1,1]}}\left(\t u_{xy[2,1]}\frac{\p h_2}{\p\t u_{xy[1,1]}}+2\,\t u_{x[1,2]}\frac{\p h_2}{\p\t u_{x[1,1]}}+\t u_{x[0,2]}\frac{\p h_2}{\p\t u_{x[0,1]}}\right)+O(1)=0, 
\end{aligned}
\end{equation} 
where $O(1)$ is a first order expression. Equating to zero the terms with variables $\t u_{xy[2,1]},\t u_{x[1,2]}$, $\t u_{x[0,2]},u_{xy[2,1]}$ and $u_{x[1,2]},$  we obtain the following equations:
$$
\frac{\p h_2}{\p\t u_{xy[1,1]}}=\frac{\p h_2}{\p\t u_{x[1,1]}}=\frac{\p h_2}{\p\t u_{x[0,1]}}=0,\quad u_{x[1,1]} \frac{\p h_2}{\p u_{x[1,1]}}=h_2,\quad u_{xy[1,1]}\frac{\p h_2}{\p u_{xy[1,1]}}=-h_2.
$$
The general solution of these equations has the following form:
\begin{equation}\label{h2}
h_2=\frac{u_{x[1,1]}}{u_{xy[1,1]}}\,h_3(\t u_{[0,0]},\t u_{y[0,1]},\t u_{y[1,1]}, u_{y[1,1]}).
\end{equation}
Taking into account this result, we can write the coefficient at the second vector derivative  $\bsy u_{xx}$ in (\ref{opreq1}) in the following form 
\begin{equation}\label{U2}
\begin{aligned}
&3\frac{u_{x[2,2]}}{u_{x[1,1]}}\,h_3+\frac{3\,a\,u_{x[1,2]}^2}{u_{x[1,1]}\,(1+a\,u_{x[1,1]})^2}\left(2\,u_{x[1,1]}\,(1+a\,u_{x[1,1]})\frac{\p h_1}{\p u_{x[1,1]}}+h_1\right)\\
&+\frac{3\,a\,u_{x[1,2]}}{1+a\,u_{x[1,1]}}\left(u_{xy[2,1]}\frac{\p h_1}{\p u_{xy[1,1]}}+\t u_{xy[2,1]}\frac{\p h_1}{\p\t u_{xy[1,1]}}+2\,\t u_{x[1,2]}\frac{\p h_1}{\p\t u_{x[1,1]}}+\t u_{x[0,2]}\frac{\p h_1}{\p\t u_{x[0,1]}}
\right)\\[2mm]
&+R=0,
\end{aligned}
\end{equation}
where $R$ is a linear function in variables $u_{xy[2,1]}, \t u_{xy[2,1]},u_{x[1,2]},\t u_{x[1,2]},\t u_{x[0,2]}$ (the explicit form of $R$ is rather cumbersome). The first term in (\ref{U2}) 
implies $h_3=0$. Taking into account (\ref{h2}), we get $h_2=0.$

It can be easily checked that if we substitute  $h_3=0$ and $a=0$ into  (\ref{U2})  then we get the contradiction  $u_{xy[1,1]}=0$ and therefore $a\ne 0$.  

Equating  the coefficients at the second order variables in (\ref{U2}) to zero, we get
$$
\frac{\p h_1}{\p u_{xy[1,1]}}=\frac{\p h_1}{\p\t u_{xy[1,1]}}=\frac{\p h_1}{\p\t u_{x[1,1]}}=\frac{\p h_1}{\p\t u_{x[0,1]}}=0,\quad 2\, u_{x[1,1]}(1+a\, u_{x[1,1]})\frac{\p h_1}{\p u_{x[1,1]}}+h_1=0.
$$
From these equations it  follows that  
$$
h_1=( u_{x[1,1]})^{-1/2}\sqrt{1+a\, u_{x[1,1]}}\,f(\t u_{[0,0]},\t u_{y[0,1]},\t u_{y[1,1]}, u_{y[1,1]}).
$$ 
Substituting this expression for $h_1$  into (\ref{U3}),  we find out that (\ref{U3}) is identically true. It turns out that equation (\ref{U2}) takes the following form:
\begin{align*}
&\frac{3\,u_{x[1,2]}}{(u_{x[1,1]})^{1/2}\sqrt{\vphantom{u^b}1+a\,u_{x[1,1]}}}\left\{a\,\t  u_{xy[1,1]}\left(\frac{\p f}{\p\t u_{y[0,1]}}+2\,f\,\frac{\p f}{\p\t u_{y[1,1]}}\frac{\sqrt{\vphantom{u'}1+a\,u_{x[1,1]}}}
{\sqrt{\vphantom{u^a}u_{x[1,1]}}}\right)\right.\\
&+u_{xy[1,1]}\left[\frac{\sqrt{\vphantom{u'}1+a\,u_{x[1,1]}}}{\sqrt{\vphantom{u^a}u_{x[1,1]}}}\left(a\,\frac{\p(f^2)}{\p u_{y[1,1]}}+1\right)- 2\,a\,\t u_{y[0,1]}\frac{\p f}{\p \t u_{y[1,1]}}
- a\,\t u_{[0,0]}\frac{\p f}{\p \t u_{y[0,1]}}\right]\\
&\left.+a\,\t  u_{x[0,1]}\left(2\,\frac{\p f}{\p\t u_{[0,0]} }+f\frac{\p f}{\p\t u_{y[0,1]}}\frac{\sqrt{\vphantom{u'}1+a\,u_{x[1,1]}}}{\sqrt{\vphantom{u^a}u_{x[1,1]}}}\right)\right\}=0.
\end{align*}
Since the function $f$ does not depend on $\t  u_{xy[1,1]},  u_{xy[1,1]},\t  u_{x[0,1]},\t u_{x[1,1]},$ we equate the coefficients at these variables to zero and find that 
$\ds f=a^{-1/2}\sqrt{b-u_{y[1,1]}}$, where $b$ is an arbitrary constant. So, we get the equation 
\begin{equation}\label{ee1}
\bsy u_{xy}=\frac{\sqrt{b-\bsy u_{y}^2}\sqrt{c+\bsy u_x^2}}{\sqrt{\bsy u_x^2}}\,\bsy u_x -(\bsy u_x,\bsy u_y)\,\bsy u,\qquad c=a^{-1}.
\end{equation}
For this equation the compatibility condition (\ref{opreq1}) is identically satisfied. If $c\ne0$ we may normalize $c$ to 1 by the scaling $x\to x\,c^{-1/2}$.  

Note that the limit of equation (\ref{A01}) as $a\to\infty $ is equation (\ref{A3}). Setting $c=0$ in (\ref{ee1}), we obtain equation
\begin{equation}\label{ee2}
\bsy u_{xy}=\sqrt{b-\bsy u_{y}^2}\,\bsy u_x -(\bsy u_x,\bsy u_y)\,\bsy u,
\end{equation}
which has $x$-symmetry (\ref{A3}). 

To complete the investigation of equation (\ref{ee1}) we have to compute its $y$-symmetry. Analyzing the list of Appendix 1, we observe that any integrable  $y$-symmetry can be 
written in the following form  
\begin{equation}\label{sym2a}
\bsy u_\tau  =  {\bsy u}_{yyy}+g_2 ( {\bsy u}_{yy}+\bsy u_y^2\,\bsy u)+g_1  {\bsy u}_y+3\,(\bsy u_y,\bsy u_{yy})\,\bsy u,
\end{equation}
where
\begin{align*}
g_2&=-\dfrac32\frac{d}{dy}\ln p_0,\\
\intertext{and}
g_1&=\frac{c_1}{p_0}\,u_{y[2,2]}+\frac{c_2}{p_0}\,\t u_{y[2,2]}+p_1\,u_{y[1,2]}^2+p_2\,\t u_{y[1,2]}^2+p_3\,\t u_{y[0,2]}^2+p _4\,u_{y[1,2]}\,\t u_{y[1,2]}\\
&+p _5\,u_{y[1,2]}\t u_{y[0,2]}+p _6\t u_{y[1,2]}\,\t u_{y[0,2]}+p_7\,u_{y[1,2]}+p_8\,\t u_{y[1,2]}+p _9 \t u_{y[0,2]}+p_{10}.\n
\end{align*}
Here  $c_1$ and $c_2$ are constants and $p_i$ are functions depending on $\t u_{[0,0]},\,\t u_{y[0,1]},\,\t u_{y[1,1]}$ and $u_{y[1,1]}$. 
It is easy to verify that this anzats is invariant with respect to the equivalence transformations. 

In the case of equation (\ref{ee1}) the functions $p_i$ can be found from 
the compatibility condition for (\ref{ee1}) and  (\ref{sym2a}). If $b\ne0$ we get the following $y$-symmetry:
\begin{equation}\label{sya2}
\bsy u_\tau=\bsy u_{yyy}+\frac{3}{2}\left(\frac{\bsy u_{yy}^2-\bsy u_y^4}{b}+\frac{(\bsy u_y,\bsy u_{yy})^2}{b\,(b-\bsy u_y^2)}+\bsy u_y^2\right)\bsy u_y+3\,(\bsy u_y,\bsy u_{yy})\,\bsy u.
\end{equation}
This equation is related to (\ref{A2}) with $a=-b^{-1}$. The rescaling $y\to y\,b^{-1/2}$ gives rise to $b=1$ and we obtain
equation (\ref{a01}), where $c=1$ or $c=0$.

If $b=0$ then $y$-symmetries of form (\ref{sym2a}) do not exist.  
However, after the rescaling $\tau\to(2/3)\, b\, \tau$   we derive from (\ref{sya2}) the following symmetry  of the first order (see Section 3.3) in the vector variables:  
\begin{equation}\label{sym1or}
\bsy u_\tau =\left({\bsy u_{yy}^2}-\frac{(\bsy u_y,\bsy u_{yy})^2}{\bsy u_y^2}-\bsy u_y^4\right)\bsy u_y.
\end{equation}
The study of the case of $x$-symmetry (\ref{A1}) is completed.

Calculations analogical to the presented above show that for each of equations (\ref{A5}) -- (\ref{A8}), (\ref{A10}), (\ref{A11}), (\ref{A13}) and (\ref{A14}) taking as $x$-symmetry the corresponding hyperbolic equation does not exist. 

Equations  (\ref{A4}) and  (\ref{A19}) generate hyperbolic equations 
\begin{equation}
\bsy u_{xy}=(\bsy u_x,\bsy u_y)\Big((p+1)\bsy u_x^{-2}\bsy u_x-\bsy u\Big)+(p-1)\bsy u_y,\qquad p=\sqrt{1-\bsy u_x^2}\,, \label{h5}  
\end{equation}
\begin{equation}\begin{array}{cc}
\ds \bsy u_{xy}=\left(\frac{\langle\bsy u,\bsy u_y \rangle}{\langle\bsy u,\bsy u\rangle}+\frac{\langle\bsy u,\bsy u\rangle \langle\bsy u_x,\bsy u_y \rangle-
\langle\bsy u,\bsy u_x \rangle\langle\bsy u,\bsy u_y \rangle}{(\mu+\langle\bsy u,\bsy u\rangle)\langle\bsy u,\bsy u\rangle}\right)\bsy u_x\\[5mm]
\ds -(\bsy u_x,\bsy u_y)\bsy u    -\frac{\mu+\langle\bsy u,\bsy u\rangle-\langle\bsy u,\bsy u_x \rangle}{\langle\bsy u,\bsy u\rangle}\bsy u_y,
\end{array} \label{h6} 
\end{equation}
where
$\mu=\sqrt{\ds\langle\bsy u,\bsy u_x\rangle^2+\langle\bsy u,\bsy u\rangle^2-\langle\bsy u,\bsy u\rangle\langle\bsy u_x,\bsy u_x \rangle},$ correspondingly. 
These equations have no $y$-symmetries of the form  (\ref{sym2a}) (see subsection 3.2).

For each of equations (\ref{A1}), (\ref{A2}), 
(\ref{A9}),  (\ref{A12}), (\ref{A15}) -- (\ref{A17}) the corresponding hyperbolic equation is completely determined by the compatibility condition (\ref{opreq1}) and we use
 (\ref{opreq2}) only to verify that a $y$-symmetry  of the form (\ref{sym2a}) exists. As a result we get  the list (\ref{a01}) -- (\ref{a16})  (see Remark 1 for details). 

In the cases (\ref{A3}) and (\ref{A18}) with $c=0$ the right hand side of (\ref{sph_eq}) contains an undetermined function of one variable while
 (\ref{opreq1}) is already satisfied.

Consider the $x$-symmetry 
$$
\bsy u_t= \bsy u_{xxx}-3\,\frac{(\bsy u_x,\bsy u_{xx})}{\bsy u_x^2}\,\bsy u_{xx}+ \frac{3}{2}\frac{\bsy u_{xx}^2}{\bsy u_x^2}\,\bsy u_x,  
$$
which is the vector Schwartz-KdV labeled (\ref{A3}) in Appendix 1. Computations show that the corresponding hyperbolic equation has the form 
\begin{equation}\label{sch1}
\bsy u_{xy}=f(\t u_{[0,0]},\t u_{y[0,1]},\t u_{y[1,1]}, u_{y[1,1]}) \, \bsy u_x-(\bsy u_x,\bsy u_y)\bsy u.
\end{equation}
For equation (\ref{sch1}) condition  (\ref{opreq1}) turns out to be equivalent to  
\begin{equation}\label{eqh1}
\begin{aligned}
&\frac{\p f}{\p \t u_{[0,0]}}=f^2\frac{\p f}{\p\t u_{y[1,1]}},\qquad \frac{\p f}{\p\t u_{y[0,1]}}=-2f\frac{\p f}{\p\t u_{y[1,1]}},\\
&2f\frac{\p f}{\p u_{y[1,1]}}+2(\t u_{[0,0]}f-\t u_{y[0,1]})\frac{\p f}{\p\t u_{y[1,1]}}+1=0.
\end{aligned}
\end{equation}
This system is compatible and therefore its general solution $f$ contains one arbitrary function of one variable.

The hyperbolic equation 
\begin{align}\label{sch2}
\bsy u_{xy}&=g(\t u_{[0,0]},\t u_{y[0,1]},\t u_{y[1,1]}, u_{y[1,1]}) \,\bsy u_x+\frac{\langle\bsy u,\bsy u_x\rangle}{\langle\bsy u,\bsy u\rangle}\bsy u_y-(\bsy u_x,\bsy u_y)\bsy u,
\end{align}
where $g$ is any solution of the compatible system
\begin{equation}\label{eqh2}
\begin{aligned}
& \frac{\p g}{\p\t u_{y[0,1]}}=-2\,g\frac{\p g}{\p\t u_{y[1,1]}},\qquad 2\,g\frac{\p g}{\p u_{y[1,1]}}+2(\t u_{[0,0]}g-\t u_{y[0,1]})\frac{\p g}{\p\t u_{y[1,1]}}+1=0,\\[2mm]
&\frac{\p g}{\p \t u_{[0,0]}}=\frac{\p g}{\p\t u_{y[1,1]}}\left(u_{y[1,1]}+g^2+\frac{g\,\t u_{y[0,1]}-\t u_{y[1,1]}}{\t u_{[0,0]}}-\frac{u_{y[1,1]}\,\t u_{y[0,1]}}{\t u_{[0,0]}\,g}\right)
+\frac{u_{y[1,1]}+g^2}{2\,\t u_{[0,0]}\,g}
\end{aligned}
\end{equation}
has integrable  $x$-symmetry  
\begin{align}\label{A180}   
\bsy u_t&=\bsy u_{xxx}+3\left(\frac{\langle\bsy u,\bsy u_x\rangle}{\langle\bsy u,\bsy u\rangle}+\frac{\langle\bsy u,\bsy u_x\rangle\langle\bsy u,\bsy u_{xx}\rangle}{\xi}-
\frac{\langle\bsy u,\bsy u\rangle\langle\bsy u_x,\bsy u_{xx}  \rangle}{\xi}\right)(\bsy u_{xx}+ \bsy u_x^2\bsy u)  \\[2mm]
&\quad+\frac{3}{2}\left(\rule[-2pt]{0pt}{2.6em}\frac{\langle\bsy u,\bsy u\rangle \langle\bsy u_{xx},\bsy u_{xx}\rangle }{\xi}-\frac{\big(\xi+\langle\bsy u,\bsy u\rangle\langle\bsy u,\bsy u_{xx}\rangle\big)^2}
{\xi\, \langle\bsy u,\bsy u\rangle^2}\right)\bsy u_x+3 (\bsy u_x,\bsy u_{xx})\bsy u,\n 
\end{align}
where $\quad \xi=\langle\bsy u,\bsy u\rangle\langle\bsy u_x,\bsy u_x\rangle -\langle\bsy u,\bsy u_x\rangle^2.$ This $x$-symmetry is nothing but equation  (\ref{A18}) with $c=0$.  
The general solution of  system (\ref{eqh2}) contains an arbitrary function of one variable.

An additional equation 
$$
\frac{\p f}{\p\t u_{y[1,1]}}\,\Big(c_2(\t u_{y[0,1]}-f\t u_{[0,0]})-c_1\,f\Big)=\frac{c_2}{2}
$$
for the function $f$ in (\ref{sch1}) can be extracted from the existence of the third order $y$-symmetry. 
This equation together with (\ref{eqh1}) allows us to find $f$. However there exists a much more economic way to verify that 
integrable hyperbolic equations with $x$-symmetries (\ref{A3}) and (\ref{A18}) appear also in some of cases (\ref{A1}), (\ref{A2}), 
(\ref{A9}),  (\ref{A12}), (\ref{A15}) -- (\ref{A17}).  Namely, it can be easily verified that neither of equations (\ref{sch1}), (\ref{sch2}) has a $y$-symmetry equivalent to (\ref{A3}) or to (\ref{A18}).
 $\square$

\rem  In this remark we indicate the symmetries of the form  (\ref{sym1}) and  (\ref{sym2}) for hyperbolic equations (\ref{a01}) -- (\ref{a16}). 

The $x$-symmetry  (\ref{sym1}) for equation (\ref{a01}) with $c=1$ is given by (\ref{A1}) with $a=1.$ The $y$-symmetry (\ref{sym2}) is defined by (\ref{A2}) with $a=-1$. 
Here and in the sequel, when we refer to Appendix 1, we assume that the replacements $x\to y$, $t\to\tau$ are performed  to obtain the $y$-symmetry .
Equation (\ref{a01}) with $c=0$ has the $x$-symmetry  (\ref{A3}) and the $y$-symmetry of the form (\ref{A2}) with $a=-1$. 
 
Equation (\ref{a12}) has $x$-symmetry (\ref{A12}) and $y$-symmetry (\ref{A18}).  

Equation (\ref{a15}) has $x$-symmetry (\ref{A15}). The form of the $y$-symmetry depends on the parameter $c$ in (\ref{a15}). If $c=0$ then the $y$-symmetry is given by (\ref{A3}).  
If $c\ne0$ then normalizing  $c$ to 1 by the scaling $y\to y\,c^{-1/2},$ we arrive at the hyperbolic equation (\ref{a15}) with the $y$-symmetry  (\ref{A9}).   

Equation (\ref{a16}) has $x$-symmetry (\ref{A16}) while the form of the $y$-symmetry depends on the values of the parameters in (\ref{a16}). If $a\ne0$  we may normalize $a$ to 1 by the 
scaling $y\to y\,a^{-1/2}$. So, we assume that $a=0$ or $a=1$. 
If  $a=1$ 
then the $y$-symmetry can be reduced to (\ref{A17}) with $c=-b$  by the transformation   $\langle \ , \rangle\to\langle \ , \rangle-b\,(\ ,\, )$.
If $a=0$ 
the transformation   $\langle \ , \rangle\to\langle \ , \rangle-b\,(\ ,\, )$ reduces the $y$-symmetry to (\ref{A180}). 
$\square$

\rem Any anisotropic equation admits  the reduction $\langle\,\cdot\,, \,\cdot \rangle \to\gamma (\,\cdot\,, \,\cdot)$ (the isotropic limit). Under this reduction we have $\langle\bsy u,\bsy u\rangle\to\gamma ,\ 
\langle\bsy u,\bsy u_x\rangle\to0,\ \langle\bsy u_x,\bsy u_x\rangle\to\gamma\bsy u_x^2$ and so on. For example,  the isotropic limit of equation (\ref{a12}) has the form
\begin{align*}
\bsy u_{xy}=\bsy u_y\,\sqrt{\vphantom{\rule{0pt}{.79em}}\gamma- \bsy u_x^2}\,\sqrt{1+c\,(\gamma \bsy u_y^{2})^{-1}}\, -(\bsy u_x,\bsy u_y)\bsy u.
\end{align*}
Performing  the scaling $x\to x\gamma ^{-1/2},\ y\to y\gamma ^{1/2}$ and replacement $x\leftrightarrow y$ we obtain  equation (\ref{a01}). The analogous computations 
show that equations (\ref{a15}) and (\ref{a16}) are also reduced to (\ref{a01}) .  $\square$

\section{Reductions of integrable hyperbolic equations}

\subsection{One-component reductions}

All equations (\ref{a01})--(\ref{a16}) are vector generalizations of different special cases for known scalar integrable equation \cite{zibsok,bz} 
\begin{equation} 
u_{xy}= P(u) \sqrt{\vphantom{u_y^2}u_x^2+1}\sqrt{k\,u_y^2+a}. \label{eqq}
\end{equation}
Here $P$ is any solution of differential equation  
\begin{equation} \label{PP}
P'^2=\lambda_1 P^4+\lambda_2 P^2+\lambda_3,  
\end{equation} 
where $\lambda_i$ are constants. 
The generic solution of this ODE is  the elliptic Jacobi sine.  

Indeed, consider the case $N=2.$ Let the scalar products be ${\bsy u}^2=u_1^2+u_2^2$ and $\langle\bsy u,\bsy u \rangle=A u_1^2+B u_2^2,\ A\ne B$. 
Using the trigonometric parametrization $\bsy u=\{\cos w ,\sin w \}$ of $S^1,$ one can rewrite equations (\ref{a01})--(\ref{a16}) in terms of the scalar variable $w$.
All thus obtained equations for $w$ have the form
$$
w_{xy}=F(\cos w)\,\sqrt{\vphantom{\rule{0pt}{.79em}}Q_1(\cos w)+c_1\,w_x^2}\,\sqrt{Q_2(\cos w)+c_2\,w_y^2}-f(\cos w, \sin w)\,w_x\,w_y,
$$
where $c_i$ are constants, $Q_i$ are polynomials, $F$ and $f$ are rational functions. 

We can reduce $f$ to zero introducing a new variable $v$ as a solution of the following equation
$$
\frac{dv}{dw}=\exp\left(\int f(w)\,dw\right)\equiv \phi (w).
$$
Let $w=\psi (v)$ be the inverse function. 
Using the relations $v_x=\phi (w)\,w_x,\ v_y=\phi (w)\,w_y,\ v_{xy}=\phi (w)\,w_{xy}+\phi' (w)\,w_x\,w_y,$ we get  
$$
v_{xy}=R(w)\sqrt{\vphantom{\rule{0pt}{.79em}}k_1+c_1\,v_x^2}\,\sqrt{k_2+c_2\,v_y^2},
$$
where $k_i$ are some constants. To prove that the function $P(v)=R(\psi (v))$  satisfies equation (\ref{PP}) one can use  
the relations
$$
R(w)=P(v),\qquad P'(v)=\frac{R'(w)}{\phi (w)}.
$$ 
 
For example, the one-component reduction of equation (\ref{a12}) is described by
$$
v_{xy}=\sqrt{\vphantom{\rule{0pt}{.79em}}\cos (2\,w)+\alpha}\,\sqrt{\vphantom{\rule{0pt}{.79em}}c_1-v_x^2}\,\sqrt{\beta +v_y^2},\qquad 
v'(w)=\frac{1}{\ds\sqrt{\vphantom{\rule{0pt}{.78em}}\cos (2\,w)+\alpha}}.
$$ 
Here the parameters $\alpha$ and $c_1$ are related to the metric coefficients by
\begin{equation}\label{mpar}
A=c_1(\alpha+1),\qquad  B=c_1(\alpha-1),\qquad c_1\ne0.
\end{equation}
and $\beta $ is defined by the relation $\beta  \,A B=c\,c_1,$ where $c$ is the parameter in (\ref{a12}). It is easy to verify that the function 
$\ds P(v)=\sqrt{\vphantom{\rule{0pt}{.78em}}\cos (2\,w)+\alpha}$ satisfies the equation  $P'^2=1-(P^2-\alpha)^2$.
So, if $c\ne0$ in  (\ref{a12}) we obtain the generic equation of the form (\ref{eqq}). 
The case $c=0$ corresponds to (\ref{eqq}) with $a=0$.  

The scalar reduction for  (\ref{a01}) with $c=1$ is the equation $w_{xy}=\sqrt{\vphantom{u_y^2}1-w_y^2}\,\sqrt{\vphantom{w_y^2}1+w_x^2},$
which is equivalent to (\ref{eqq}) with $P=1$.  The reduction of equation (\ref{a01}) with $c=0$  is given by $w_{xy}=w_x\sqrt{1-w_y^2}.$  

The reduction for equation  (\ref{a15}) has the form
$$
v_{xy}=P(v)\sqrt{c_1^{-1}v_y^2+c}\,\sqrt{c_1(1-\alpha^2)v_x^2+1}, 
$$
where $\alpha$ and $c_1$ are given in (\ref{mpar}) and $P$ satisfies the equation  $P'^2=P^4-(1-\alpha\,P^2)^2$.

Equation (\ref{a16}) with $b\ne0$ is reduced to 
\begin{equation}\label{ss6}
v_{xy}=P(v)\,\sqrt{1+c_1^2(1-\alpha ^2)\,v_x^2}\,\sqrt{a\,(\beta^2-1)^{-1}+v_y^2},
\end{equation}
where $P'^2=(P^2-1)^2-(\beta -\alpha P^2)^2$.  The parameters in (\ref{ss6}) are defined by (\ref{mpar}) and by the relation
$b=c_1(\beta -\alpha ).$  If $b=0$, then the reduction is given by  
$$
v_{xy}=\sqrt{\vphantom{\rule{0pt}{.78em}}1+c_1^2(1-\alpha ^2)\,v_x^2}\,\sqrt{a\,(\alpha ^2-1)^{-1}+v_y^2}.
$$
 
Thus, for generic values of parameters  equations (\ref{a12}), (\ref{a15}) and  (\ref{a16}) are reduced to   equation   (\ref{sn})

\subsection{Integrals for degenerations}
It is known that there exist two different classes of integrable equations (\ref{hyper}): the Liouville and the sin-Gordon type equations. Very often equations of the Liouville type can be obtained from some sin-Gordon type equations when a parameter in the latter equation tends to zero. For example, if we set $c_2=0$ in the sh-Gordon equation $u_{xy}=c_1 \exp{(u)}+c_2 \exp{(-u)},$ then we get
(up to a shift of $u$) the Liouville equation $u_{xy}=\exp (u).$  

The sin-Gordon type equations are integrable by the inverse scattering method (see for example \cite{ZM}) while the integrating of the equations of the Liouville type (also 
named as the Darboux integrable equations) can be reduced to solving of ordinary differential equations. This reduction to ODEs turns out to be possible because of the existence of so-called $x$ and $y$-integrals.

A function $I$ depending on   $ u,  u_y, u_{yy},\dots,\ u_{y}^{(n)}$ is called the $n$-th order $x$-integral of hyperbolic equation  (\ref{hyper}) if the identity 
$$
\frac{d }{d x}\,I=0
$$
is valid for any solution $u(x,y)$ of (\ref{hyper}). For example, the function $I=u_{yy}-\frac{1}{2} u_y^2$ is an $x$-integral for the Liouville equation. 
The  $y$-integrals are functions $J(u,  u_x, u_{xx},\dots)$ such that $\frac{d }{d y}\,J=0$.

For vector equations of the form (\ref{h1}) the scalar $x$--integrals are functions $I$ depending on scalar products of vectors $\bsy u,\bsy u_y,\bsy u_{yy},\dots \,.$ 
Moreover, vectorial integrals exist for some vector  hyperbolic equations.  For instance,  equation (\ref{h5}) has the following vectorial $y$-integral
\begin{equation*}
\bsy J_1=\frac{\bsy u_x}{\sqrt{1-p}}+\sqrt{1-p}\ \bsy u.  
\end{equation*}
It is easy to verify that $\bsy J_1^2=2.$
Setting $\bsy J_1=\bsy v(x),$ where $\bsy v$ is an arbitrary vector such that  $ \bsy v^2=2,$ we obtain the following ordinary  differential  equation
\begin{equation*}
\bsy u_x=(\bsy u,\bsy v)\big[\bsy v-(\bsy u,\bsy v)\,\bsy u\big]
\end{equation*}
equivalent to  (\ref{h5}). 
The $y$-integral of equation (\ref{h6}) has the form
\begin{equation*}
\bsy J_2=\frac{\bsy u_x}{\sqrt{\mu+\langle\bsy u,\bsy u\rangle}}- 
\frac{\mu+\langle\bsy u,\bsy u\rangle+\langle\bsy u,\bsy u_x\rangle}{\langle\bsy u,\bsy u\rangle\sqrt{\mu+\langle\bsy u,\bsy u\rangle}}\,\bsy u
\end{equation*}
and the equation is equivalent to
 \begin{equation*}
\bsy u_x=\langle\bsy u,\bsy w\rangle\big[(\bsy u,\bsy w)\,\bsy u-\bsy w\big], 
\end{equation*}
where $\bsy J_2=\bsy w(x),\quad \langle\bsy w, \bsy w\rangle=2.$ 

Equations (\ref{sch1}) and  (\ref{sch2}) with arbitrary functions $f$ and $g$ have the following vectorial $x$-integrals 
\begin{equation}
\bsy I_1=\bsy u_{y}-f\bsy u  \label{a3int}
\end{equation}
and 
\begin{equation}
\bsy I_2=\frac{\bsy u_{y}-g\bsy u}{\sqrt{\langle\bsy u,\bsy u\rangle}} \label{a18int}
\end{equation}
correspondingly. Setting $\bsy I_1=\bsy v(y)$,  where $\bsy v$ is the arbitrary vector function, we get  the  ordinary  differential  equation
$$
\bsy u_{y}=\bsy v(y)- (\bsy u,\bsy v)\bsy u 
$$
equivalent to  (\ref{sch1}). Equation (\ref{sch2}) is equivalent to 
$$
\bsy u_{y}=\sqrt{\langle\bsy u,\bsy u\rangle}\ \big[\bsy w-(\bsy u,\bsy w)\,\bsy u\big]
$$
where $\bsy I_2=\bsy w(y)$.

For particular values of the parameters equations (\ref{a01})  -- (\ref{a16}) are equivalent to special cases of (\ref{sch1}) or to (\ref{sch2}) and therefore have vectorial integrals.  
Equation  (\ref{a01}) with $c=0$ has the vectorial $x$-integral  (\ref{a3int}) with $f=\sqrt{1-\bsy u_y^2}$.
Equation (\ref{a15})  with $c=0$  is related to equation (\ref{sch1})  by  $x\leftrightarrow y$ and therefore
 equation (\ref{a15}) with $c=0$ has the $y$-integral   $\bsy u_{x}-f\bsy u$, where
$$
f=\frac{\langle\bsy u,\bsy u_x \rangle+\sqrt{\vphantom{\rule{0pt}{.78em}}\xi}}{\langle\bsy u,\bsy u \rangle}
$$
and $\xi$ is defined in  (\ref{a15}). 
Equation (\ref{a12}) with $c=0$ has the form  (\ref{sch2}) up to  $x\leftrightarrow y$ and so this equation has the  $y$-integral given by (\ref{a18int}) with $x\leftrightarrow y$.
Equation (\ref{a16}) with $a=0$  is related to (\ref{sch2}) by   $x\leftrightarrow y$
and by the transformation $\langle\, , \rangle\to\langle\, , \rangle+b\,(\, , )$. Therefore the form of the $y$-integral is slightly differ from (\ref{a18int}):
$$
\bsy J= \frac{1}{\sqrt{\eta }}\left(\bsy u_x-\frac{\langle\bsy u,\bsy u_x\rangle+\sqrt{ \psi}}{\langle\bsy u,\bsy u\rangle}\,\bsy u\right).
$$

There exists another degeneration process for equations (\ref{a01}) -- (\ref{a16}) that yields hyperbolic equations with {\it scalar} integrals. For example, consider equation  
(\ref{ee1}), which can be obtained from (\ref{a01}) 
by the rescaling  $y\to\sqrt{b}\,y$. If $b=0$ in (\ref{ee1}) we get the equation  
$$
\bsy u_{xy}=i\sqrt{\bsy u_y^2}\sqrt{\vphantom{\bsy u_y^2}1+c\,\bsy u_x^{-2}}\ \bsy u_x-(\bsy u_x,\bsy u_y)\bsy u.
$$
This equation has the scalar $x$-integral
$$
I={\bsy u_{yy}^2}-\frac{(\bsy u_y,\bsy u_{yy})^2}{\bsy u_y^2}-\bsy u_y^4
$$
for any $c$. The third order symmetry of equation (\ref{ee1}) degenerates to a first order symmetry of the form  $\bsy u_\tau  =I\,\bsy u_y$ (see formulas (\ref{sya2}) and (\ref{sym1or})).

For equations (\ref{a12}), (\ref{a15}) we should perform the rescaling $x\to\sqrt{b}\,x$ and put $b=0$ to obtain the degenerations.  In the case of equation (\ref{a16}) we apply the transformation $x\to\sqrt{c}\,x$ and put $c=0$.
 Equations obtained by this procedure have scalar $y$-integrals. 

 For equations  (\ref{a12}) -- (\ref{a16})  the third order symmetry degenerates to a first order symmetry of the form  $\bsy u_t =J\,\bsy u_x,$ where $J$ is the corresponding $y$-integral.

\section{Auto-B\"acklund transformations}

Consider two copies 
\begin{align}\label{sph_eqs1}
{\bsy u}_{xy}&=h_1( \bsy u) {\bsy u}_x+h_2( \bsy u)  {\bsy u}_y-(\bsy u_x,\bsy u_{y}) {\bsy u},\\ 
\intertext{and}
{\bsy v}_{xy}&=h_1( \bsy v)  {\bsy v}_x+h_2( \bsy v)  {\bsy v}_y -(\bsy v_x,\bsy v_{y}) {\bsy v}.\label{sph_eqs2}
\end{align}
of the same hyperbolic equation (\ref{sph_eq}).
The first order auto-B\"acklund transformation (BT) for equation (\ref{sph_eq}) is a pair of the ordinary differential equations of the form
\begin{equation}\label{BT}
\bsy u_x=f_0\bsy u+f_1\bsy v+f_2\bsy v_x,\qquad \bsy u_y=g_0\bsy u+g_1\bsy v+g_2\bsy v_y,
\end{equation}
compatible with  (\ref{sph_eqs1}), (\ref{sph_eqs2}). The functions $f_i$ in (\ref{BT}) depend on scalar products of the vectors $\bsy u,\bsy v$ and $\bsy v_x$, 
whereas $g_i$ are functions in scalar products of the vectors $\bsy u,\bsy v$ and $\bsy v_y$. 
Equations (\ref{BT}) are called the {\it $x$-component} and  {\it $y$-component} of the BT correspondingly.

It follows from the relations $\bsy u^2=1,\ (\bsy u,\bsy u_x)=(\bsy u,\bsy u_y)=0$ that
$f_0=-f_1(\bsy u,\bsy v)-f_2(\bsy u,\bsy v_x),\ g_0=-g_1(\bsy u,\bsy v)-g_2(\bsy u,\bsy v_y)$.

If a solution $\bsy u (x,y)$ of equation (\ref{h1}) is given, then one can find another solution $\bsy v (x,y)$ of the same equation by solving ODEs (\ref{BT}).
Suppose that the coefficients of (\ref{BT}) depend on arbitrary ``spectral'' parameter $\lambda$. Then starting from a simple solution of (\ref{h1}) and applying (\ref{BT}) 
several times, we can  construct a multi-parametric family of solutions. The so-called multi-solitonic solutions for classical integrable systems can be found in this way.  
By that reason the existence of BT with the spectral parameter can be regarded as a strong indication of the complete integrability. 

To justify the integrability of equations (\ref{a01}) -- (\ref{a16}) we present  auto-B\"acklund transfor\-mations with the spectral parameter for them.  Equation (\ref{a01}) has the following BT:
\begin{equation*}
\begin{aligned}
&\bsy u_x=f\Big[(\bsy u,\bsy v_x)(\bsy u+\bsy v)-h\,\bsy v_x\Big],\qquad h=(\bsy u,\bsy v)+1, \\
&\bsy u_y=\left(\frac{(\bsy u,\bsy v_y)(h+\lambda )}{h\,(\lambda +1)}+
(\bsy u,\bsy v)\,g\right)\bsy u+\left(\frac{\lambda (\bsy u,\bsy v_y)}{h\,(\lambda +1)}-g\right)\bsy v-\bsy v_y,
\end{aligned}
\end{equation*}
where
$$
f=\frac{1}{\lambda }+\frac{1}{h}+\frac{1}{\lambda }\sqrt{1+2\lambda \,h^{-1}}\sqrt{1+c\,\bsy v_x^{-2}},\qquad g=\frac{1}{\lambda +1}\sqrt{1+2\lambda \,h^{-1}} \sqrt{1-\bsy v_y^2} .
$$
 
The BT for equation (\ref{a12}) is given by
\begin{equation*}
\begin{aligned}
&\bsy u_x=f\left(\bsy v_x-\bsy u\,(\bsy u,\bsy v_x)+\frac{\lambda\,(\bsy u,\bsy v_x)+h\sqrt{\langle\bsy v,\bsy v\rangle -\bsy v_x^2}}
{\langle\bsy u,\bsy v\rangle-f\langle\bsy v,\bsy v\rangle}\,\big(\bsy v-\bsy u\,(\bsy u,\bsy v)\big)\right), \\
&\bsy u_y=\frac{1}{\lambda }\Big(q+h\,\sqrt{1+c\,\langle\bsy v,\bsy v\rangle\,\psi ^{-1}}\Big)\Big(\bsy v_y-\bsy u\,(\bsy u,\bsy v_y)-g\,\big(\bsy v-\bsy u\,(\bsy u,\bsy v)\big)\Big),
\end{aligned}
\end{equation*}
where 
$$
\begin{aligned}
&f=\sqrt{\frac{\langle\bsy u,\bsy u\rangle}{\langle\bsy v,\bsy v\rangle}}\,,\qquad g=\frac{\langle\bsy u,\bsy v_y\rangle-f\langle\bsy v,\bsy v_y\rangle}
{\langle\bsy u,\bsy v\rangle-f\langle\bsy v,\bsy v\rangle}\,,\qquad h=\sqrt{q^2-\lambda ^2},\\
&q=\langle\bsy u,\bsy v\rangle-f\langle\bsy v,\bsy v\rangle+\lambda (\bsy u,\bsy v),\qquad \psi=\langle\bsy v,\bsy v\rangle\langle\bsy v_y,\bsy v_y\rangle-\langle\bsy v,\bsy v_y\rangle^2.
\end{aligned}
$$

The B\"acklund  transformation for equation (\ref{a15}) has the form
\begin{equation*}
\begin{aligned}
&\bsy u_x=(\bsy u,\bsy v_x)\,\bsy u-g\,\big(\bsy v-\bsy u\,(\bsy u,\bsy v)\big)-\bsy v_x,  \\[2mm]
&\bsy u_y=\frac{q+h\,\sqrt{1+c\,\langle\bsy v,\bsy v  \rangle\,\bsy v _y^{-2}}}{\langle\bsy v,\bsy v  \rangle}\left(\frac{(\bsy u,\bsy v_y)}{(\bsy u,\bsy v)+1}(\bsy u +\bsy v)-\bsy v_y\right),
\end{aligned}
\end{equation*}
where
$$
\begin{aligned}
&g=\frac{h\,\sqrt{\,\phi\,}+\langle\bsy v,\bsy v_x  \rangle\langle\bsy u,\bsy v  \rangle-\langle\bsy u,\bsy v_x  \rangle\langle\bsy v,\bsy v  \rangle}
{\lambda\,\langle\bsy v,\bsy v  \rangle \big((\bsy u,\bsy v)+1\big)}-\frac{\langle\bsy v,\bsy v_x  \rangle}{\langle\bsy v,\bsy v  \rangle},
\qquad h=\sqrt{q^2-\langle\bsy u,\bsy u \rangle\langle\bsy v,\bsy v  \rangle},\\
&q=\lambda\, \big((\bsy u,\bsy v)+1\big)-\langle\bsy u,\bsy v  \rangle, \qquad 
\phi=\langle\bsy v,\bsy v_x  \rangle^2+\langle\bsy v,\bsy v  \rangle(1-\langle\bsy v_x,\bsy v_x  \rangle).
\end{aligned}
$$

The B\"acklund  transformation for equation (\ref{a16}) can be written in the following form:
\begin{equation*}
\begin{aligned}
&\bsy u_x=f\,\left(\bsy v_x -\bsy u\,(\bsy u,\bsy v_x)+\frac{h\,\sqrt{\,\zeta\, }+q\,\langle\bsy v,\bsy v_x \rangle-\langle\bsy u,\bsy v_x \rangle\langle\bsy v,\bsy v\rangle}
{\langle\bsy v,\bsy v\rangle\,\big(\langle\bsy u,\bsy v\rangle-q\big)}\,\big(\bsy v -\bsy u\,(\bsy u,\bsy v )\big)\right),\\
&\bsy u_y=\frac{q-h\,\sqrt{1+a\,\langle\bsy v,\bsy v\rangle\,\theta\, \nu^{-1}}}{\langle\bsy v,\bsy v\rangle}\Big(\bsy v_y -\bsy u\,(\bsy u,\bsy v_y)-g\,\big(\bsy v -\bsy u\,(\bsy u,\bsy v)\big)
\Big),
\end{aligned}
\end{equation*}
where 
$$
\begin{aligned}
&f=\sqrt{\frac{\eta }{\theta }}\,,\qquad \eta =\langle\bsy u,\bsy u\rangle+b,\qquad  \theta =\langle\bsy v,\bsy v\rangle+b,\qquad h=\sqrt{q^2-\langle\bsy u,\bsy u\rangle\langle\bsy v,\bsy v\rangle}\,, \\[2mm] 
&q=\langle\bsy u,\bsy v\rangle+\lambda\,\big(\langle\bsy u,\bsy v\rangle +b\,(\bsy u,\bsy v)-\theta \,f\big), \qquad \zeta =\langle\bsy v,\bsy v\rangle\big(\theta -\langle\bsy v_x,\bsy v_x\rangle\big)+\langle\bsy v,\bsy v_x\rangle^2, \\[2mm] 
&\nu =\theta \big(\langle\bsy v_y,\bsy v_y\rangle+b\,\bsy v_y^2\big)-\langle\bsy v,\bsy v_y\rangle^2,\qquad
g=\frac{\langle\bsy u,\bsy v_y\rangle+b\,(\bsy u,\bsy v_y)-f\,\langle\bsy v,\bsy v_y\rangle}{\langle\bsy u,\bsy v\rangle+b\,(\bsy u,\bsy v)-\theta \,f}.
\end{aligned}
$$ 

\section*{Appendix 1. List of evolution equations on the sphere \\ that have infinitely many conservation laws}
I. Isotropic equations 
\begin{align}
&\bsy u_t= \bsy u_{xxx}-3\,\frac{(\bsy u_x,\bsy u_{xx})}{\bsy u_x^2}\,\bsy u_{xx}+\frac{3}{2}\,\biggl(\frac{\bsy u_{xx}^2}{\bsy u_x^2}+ \tag{A.1}
\frac{(\bsy u_x,\bsy u_{xx})^2}{\bsy u_x^4\,(1+a\,\bsy u_x^2)}\biggr)\,\bsy u_x,  \label{A1}\\[1mm]
&\bsy u_t=\bsy u_{xxx}+\frac{3}{2}\biggl(\frac{a^2\,(\bsy u_x,\bsy u_{xx})^2}{1+a\,\bsy u_x^2}-a\,(\bsy u_{xx}^2-\bsy u_x^4)+\bsy u_x^2\biggr)\,\bsy u_x+3\,(\bsy u_x,\bsy u_{xx})\,\bsy u,   \tag{A.2}
\label{A2} 
\end{align}
\begin{align}
&\bsy u_t= \bsy u_{xxx}-3\,\frac{(\bsy u_x,\bsy u_{xx})}{\bsy u_x^2}\,\bsy u_{xx}+ \frac{3}{2}\frac{\bsy u_{xx}^2}{\bsy u_x^2}\,\bsy u_x,  \tag{A.3}   \label{A3} \\[1mm]
&\begin{aligned}
\bsy u_t&=\bsy u_{xxx}-3\,{\frac {\left(p+1\right )\, (\bsy u_x,\bsy u_{xx})}{2\,p\,\bsy u_x^2}}\,\bsy u_{xx}+3\,{\frac {\left(p-1\right )\,(\bsy u_x,\bsy u_{xx})}{2\,p}} \, \bsy u\\[2mm]
&+\frac{3}{2}\left ({\frac {\left (p+1\right )\bsy u_{xx}^2}{\bsy u_x^2}}-{\frac {\left(p+1\right )a\,{(\bsy u_x,\bsy u_{xx})}^{2}}{{p}^{2}\bsy u_x^2}}+\bsy u_x^2\left(1-p\right )\right )\bsy u_x,
\end{aligned}                      \tag{A.4}          \label{A4}
\end{align}
where $a$ and $c$ are arbitrary constants, $p=\sqrt{\mathstrut 1+a\,\bsy u_x^2} $. If $a=0$ and $p=\pm 1$ equation (\ref{A4}) shrinks to
\begin{align}
\bsy u_t&=\bsy u_{xxx}+3\,\bsy u_x^2\, \bsy u_x +3\,(\bsy u_x,\bsy u_{xx}) \,\bsy u.  \tag{A.5}   \label{A5}\\
\bsy u_t&=\displaystyle{ \bsy u_{xxx}-3\,\frac{(\bsy u_x,\bsy u_{xx})}{\bsy u_x^2}\,\bsy u_{xx}+ 3\,\frac{\bsy u_{xx}^2}{\bsy u_x^2}\,\bsy u_x. }  \tag{A.6}    \label{A6}
\end{align}

II. Anisotropic equations
\begin{align}\label{A7}
&\bsy u_t=\bsy u_{xxx}+\frac{3}{2}\,\Big(\bsy u_x^2+\langle\bsy u,\bsy u\rangle\Big)\,\bsy u_{x}+3\,(\bsy u_x,\bsy u_{xx})\,\bsy u,  \tag{A.7}   \\[2mm]
\label{A8}
&\bsy u_t=\bsy u_{xxx} - 3\,\frac{(\bsy u_x,\bsy u_{xx})}{\bsy u_x^2}\,\bsy u_{xx}+\frac{3}{2}\biggl(\frac{\bsy u_{xx}^2}{\bsy u_x^2}+\frac{(\bsy u_x,\bsy u_{xx})^2}{\bsy u_x^4}+
\frac{\langle\bsy u_x,\bsy u_x\rangle}{\bsy u_x^2}\biggr)\,\bsy u_{x},  \tag{A.8}  \\[2mm]
\label{A9} \tag{A.9}  
&\bsy u_t=\bsy u_{xxx}-3\frac{(\bsy u_x,\bsy u_{xx})}{\bsy u_x^2}\,\bsy u_{xx} +\frac{3}{2}\biggl(\frac{\bsy u_{xx}^2}{\bsy u_x^2}+\frac{(\bsy u_x,\bsy u_{xx})^2}{\bsy u_x^4}\\[2mm]
&\hspace{5cm}-\frac{\big(\langle\bsy u,\bsy u_x\rangle +(\bsy u_x,\bsy u_{xx})\big)^2}{\big(\bsy u_x^2+ \langle\bsy u,\bsy u\rangle\big)\,\bsy u_x^2}
+\frac{\langle\bsy u_x,\bsy u_x\rangle}{\bsy u_x^2}\biggr)\,\bsy u_{x}, \n\\[2mm]
\label{A10}  \tag{A.10}  
&\bsy u_t=\bsy u_{xxx}-3\frac{\langle\bsy u,\bsy u_x\rangle }{\langle\bsy u,\bsy u\rangle} \bsy u_{xx} 
-3\left(\frac{2\langle\bsy u,\bsy u_{xx}\rangle +\langle\bsy u_x,\bsy u_x\rangle+a}{2\langle\bsy u,\bsy u\rangle}-\frac{5}{2}\frac{\langle\bsy u,\bsy u_x\rangle^2 }{\langle\bsy u,\bsy u\rangle^2}\right)\bsy u_{x} \\
&\hspace{5cm} +3 \left((\bsy u_x,\bsy u_{xx}) -\frac{\langle\bsy u,\bsy u_x\rangle }{\langle\bsy u,\bsy u\rangle}\bsy u_x^2\right)\bsy u , \n \\[2mm]
\label{A11} \tag{A.11}  
&\bsy u_t=\bsy u_{xxx}-3\frac{\langle\bsy u,\bsy u_x\rangle }{\langle\bsy u,\bsy u\rangle} \bsy u_{xx}
-3\left(\frac{\langle\bsy u,\bsy u_{xx}\rangle  }{\langle\bsy u,\bsy u\rangle}-2\frac{ \langle\bsy u,\bsy u_x\rangle^2 }{\langle\bsy u,\bsy u\rangle^2}\right)\bsy u_{x}\\[2mm]
&\hspace{5cm}+3 \left((\bsy u_x,\bsy u_{xx}) -\frac{\langle\bsy u,\bsy u_x\rangle }{\langle\bsy u,\bsy u\rangle}\bsy u_x^2\right)\bsy u , \n \\[2mm]
%
%
\label{A12} \tag{A.12}  
&\bsy u_t= \bsy u_{xxx}- 3\frac{\langle\bsy u,\bsy u_x\rangle }{\langle\bsy u,\bsy u\rangle}\big( \bsy u_{xx}+\bsy u_x^2\bsy u \big) 
+\frac{3}{2}\left(\frac {\bsy u_{xx}^2}{\langle\bsy u,\bsy u\rangle}+\frac {\big((\bsy u_x,\bsy u_{xx})-\langle\bsy u,\bsy u_x\rangle\big )^{2}}
{\langle\bsy u,\bsy u\rangle\big(\langle\bsy u,\bsy u\rangle-\bsy u_x^2\big)}
 \right. \\[2mm]
&\qquad\left. -\frac {\big(\langle\bsy u,\bsy u\rangle- \bsy u_x^2\big)^{2}}{\langle\bsy u,\bsy u\rangle}+\frac{\langle\bsy u,\bsy u_x\rangle ^{2}}{\langle\bsy u,\bsy u\rangle^{2}}
-\frac {\langle\bsy u_x,\bsy u_x\rangle}{\langle\bsy u,\bsy u\rangle}\right) \bsy u_{x}+3 (\bsy u_x,\bsy u_{xx})\bsy u, \n \\[4mm]
%
%
\label{A13} \tag{A.13}  
&\bsy u_t=\bsy u_{xxx}+3\left( \frac{\langle\bsy u,\bsy u_x\rangle\langle\bsy u,\bsy u_{xx}\rangle  }{\xi}-\frac{\langle\bsy u_x,\bsy u_{xx}\rangle \langle\bsy u,\bsy u\rangle }{\xi}
+\frac{\langle\bsy u,\bsy u_x\rangle }{\langle\bsy u,\bsy u\rangle} \right)\big( \bsy u_{xx}+\bsy u_x^2\bsy u\big) \\[2mm]
&\quad +\frac{3}{2\xi^2\langle\bsy u,\bsy u\rangle^2 } \bigg( \Big(\langle\bsy u,\bsy u\rangle^2\langle\bsy u_x,\bsy u_{xx}\rangle -2\xi \langle\bsy u,\bsy u_x\rangle
-\langle\bsy u,\bsy u\rangle\langle\bsy u,\bsy u_x\rangle\langle\bsy u,\bsy u_{xx}\rangle \Big )^2 \n \\[2mm]
&\quad\qquad +  \langle\bsy u,\bsy u\rangle^3\langle\bsy u_{xx},\bsy u_{xx}\rangle \xi-\xi (\xi+\langle\bsy u,\bsy u_{xx}\rangle \langle\bsy u,\bsy u\rangle )^2\bigg)\bsy u_{x}
+3(\bsy u_x,\bsy u_{xx})\bsy u\n\\[2mm]
&\quad\qquad -a\frac{\langle\bsy u,\bsy u\rangle^2\bsy u_x^2+\langle\bsy u,\bsy u_x\rangle^2}{\langle\bsy u,\bsy u\rangle\xi}\bsy u_{x}, \qquad 
 \xi=\langle\bsy u,\bsy u\rangle \langle\bsy u_x,\bsy u_x\rangle-\langle\bsy u,\bsy u_x\rangle^2,\n 
\end{align}
\begin{align}
\label{A14} \tag{A.14}  
&\bsy u_t=\bsy u_{xxx}+3\left( \frac{\langle\bsy u,\bsy u_x\rangle\langle\bsy u,\bsy u_{xx}\rangle  }{\xi}-\frac{\langle\bsy u_x,\bsy u_{xx}\rangle \langle\bsy u,\bsy u\rangle }{\xi}
+\frac{\langle\bsy u,\bsy u_x\rangle }{\langle\bsy u,\bsy u\rangle} \right)\left( \bsy u_{xx}+\bsy u_x^2\bsy u\right) \\[2mm]
&\quad +\frac{3}{\xi } \left(\rule[-2pt]{0pt}{2.2em} \langle\bsy u,\bsy u\rangle\langle\bsy u_{xx},\bsy u_{xx}\rangle -2\langle\bsy u,\bsy u_x\rangle \langle\bsy u_x,\bsy u_{xx}\rangle \vphantom{\int}\right. \n \\
&\qquad\qquad\left. -\frac{\Big (\langle\bsy u,\bsy u_{xx}\rangle \langle\bsy u,\bsy u\rangle -2\langle\bsy u,\bsy u_x\rangle^2\Big)\Big(\xi+
\langle\bsy u,\bsy u_{xx}\rangle \langle\bsy u,\bsy u\rangle\Big)}{\langle\bsy u,\bsy u\rangle^2} \right)\bsy u_{x}  \n\\[2mm]
 &\qquad+3(\bsy u_x,\bsy u_{xx})\bsy u , \qquad \xi=\langle\bsy u,\bsy u\rangle \langle\bsy u_x,\bsy u_x\rangle-\langle\bsy u,\bsy u_x\rangle^2,\n  \\[4mm]
%
\label{A15} \tag{A.15}  
&\bsy u_t=\bsy u_{xxx} +\frac{3}{2}\left(\langle\bsy u_{xx},\bsy u_{xx}\rangle-\frac{\big (\langle\bsy u,\bsy u_{xx}\rangle +1\big)^2 }{\langle\bsy u,\bsy u\rangle}\right)\bsy u_{x}
+ 3\,(\bsy u_x,\bsy u_{xx}) \bsy u 
\\[2mm]
&\quad +\frac{3}{2}\left(\frac{\Big(\langle\bsy u,\bsy u\rangle\langle\bsy u_x,\bsy u_{xx}\rangle -\langle\bsy u,\bsy u_x\rangle\big(1
+\langle\bsy u,\bsy u_{xx}\rangle\big )\Big)^2 }{\xi\,\langle\bsy u,\bsy u\rangle} -  \bsy u_x^2\right)\bsy u_{x},
\n\\[2mm]
&\quad\qquad \xi=\langle\bsy u,\bsy u\rangle\big(1-\langle\bsy u_x,\bsy u_x\rangle\big)+\langle\bsy u,\bsy u_x\rangle^2, \n 
\end{align}
\begin{align}
\label{A16} \tag{A.16}  
&\bsy u_t=\bsy u_{xxx}-3\frac{\langle\bsy u,\bsy u_x\rangle}{\eta}(\bsy u_{xx}+\bsy u_x^2 \bsy u) +\frac{3}{2}\left(\frac{\langle\bsy u_{xx},\bsy u_{xx}\rangle  }{\eta}
+\frac{\xi }{\eta\,\langle\bsy u,\bsy u\rangle}\right)\bsy u_{x}\\[2mm]
&\quad +\frac{3}{2}\left(\frac{\Big(\langle\bsy u,\bsy u\rangle\langle\bsy u_x,\bsy u_{xx}\rangle 
-\langle\bsy u,\bsy u_x\rangle\big(\langle\bsy u,\bsy u_{xx}\rangle+2\,\langle\bsy u,\bsy u\rangle+b\big )\Big)^2 }{\eta\,\xi\,\langle\bsy u,\bsy u\rangle}\right. \n \\[2mm]
&\qquad\qquad  -\frac{\big(\langle\bsy u,\bsy u_{xx}\rangle +\eta\big)^2 }{\eta\,\langle\bsy u,\bsy u\rangle}
-b \frac{\langle\bsy u,\bsy u_x\rangle^2}{\eta^2\,\langle\bsy u,\bsy u\rangle}-b \frac{ \bsy u_x^2 }{\eta}\Bigg)\bsy u_{x}
+ 3\,(\bsy u_x,\bsy u_{xx}) \bsy u,  \n\\[2mm]
&\quad\qquad \eta=\langle\bsy u,\bsy u\rangle+b,\quad \xi=\langle\bsy u,\bsy u\rangle\big(\eta-\langle\bsy u_x,\bsy u_x\rangle\big)+\langle\bsy u,\bsy u_x\rangle^2, \n 
\end{align}
\begin{align}
\label{A17}  \tag{A.17}  
&\bsy u_t=\bsy u_{xxx}+3\left( \frac{\langle\bsy u,\bsy u_x\rangle}{\langle\bsy u,\bsy u\rangle }+\frac{\langle\bsy u,\bsy u_x\rangle\langle\bsy u,\bsy u_{xx}\rangle  }{\xi }
-\frac{\langle\bsy u,\bsy u\rangle\langle\bsy u_x,\bsy u_{xx}\rangle  }{\xi }\right) \big(\bsy u_{xx}+\bsy u_x^2\bsy u\big) \\[2mm]
&\quad +\frac{3}{2}\frac{ \Big(  \langle\bsy u,\bsy u\rangle^2\big(\xi\,\langle\bsy u,\bsy u_x\rangle-\eta\,\langle\bsy u_x,\bsy u_{xx}\rangle\big )
+\eta\,\langle\bsy u,\bsy u_x\rangle\big(\langle\bsy u,\bsy u\rangle\langle\bsy u,\bsy u_{xx}\rangle +\xi \big) \Big )^2 }{\eta\, \langle\bsy u,\bsy u\rangle^2(\xi+\eta)\,\xi^2 }\bsy u_{x}, \n\\[2mm]
&\quad +\frac{3}{2}\left(\frac{\langle\bsy u,\bsy u\rangle\langle\bsy u_{xx},\bsy u_{xx}\rangle  }{\xi}
-\frac{\big(\langle\bsy u,\bsy u\rangle\langle\bsy u,\bsy u_{xx}\rangle +\xi\big)^2 }{\langle\bsy u,\bsy u\rangle^2\,\xi }\right)\bsy u_{x} \n \\[2mm]
&\quad +\frac{3}{2}\,c\,\langle\bsy u,\bsy u\rangle\frac{\bsy u_x^2\,\eta+\langle\bsy u,\bsy u_x\rangle^2 }{\eta\,\xi}\bsy u_{x} +3(\bsy u_x,\bsy u_{xx})\bsy u \n \\[2mm]
&\quad \quad  \xi=\langle\bsy u,\bsy u\rangle \langle\bsy u_x,\bsy u_x\rangle-\langle\bsy u,\bsy u_x\rangle^2,\quad\eta=\big(\langle\bsy u,\bsy u\rangle+c\big)\langle\bsy u,\bsy u\rangle,\n 
\end{align}
\begin{align}\label{A18}  \tag{A.18}
\bsy u_t&=\bsy u_{xxx}+3\left(\frac{\langle\bsy u,\bsy u_x\rangle}{\langle\bsy u,\bsy u\rangle}+\frac{\langle\bsy u,\bsy u_x\rangle\langle\bsy u,\bsy u_{xx}\rangle}{\xi}-
\frac{\langle\bsy u,\bsy u\rangle\langle\bsy u_x,\bsy u_{xx}  \rangle}{\xi}\right)(\bsy u_{xx}+ \bsy u_x^2\bsy u)\n\\[2mm]
&+\frac{3}{2}\left(\rule[-2pt]{0pt}{2.6em}\frac{\langle\bsy u,\bsy u\rangle \langle\bsy u_{xx},\bsy u_{xx}\rangle }{\xi}-\frac{\big(\xi+\langle\bsy u,\bsy u\rangle\langle\bsy u,\bsy u_{xx}\rangle\big)^2}
{\xi\, \langle\bsy u,\bsy u\rangle^2}+c\,\langle\bsy u,\bsy u\rangle\frac{\bsy u_x^2}{\xi}\right.\n \\[2mm]
&\left.+c\,\frac{\Big(\langle\bsy u,\bsy u_x\rangle\big(\xi+\langle\bsy u,\bsy u\rangle\langle\bsy u,\bsy u_{xx}\rangle\big)-\langle\bsy u,\bsy u\rangle^2\langle\bsy u_x,\bsy u_{xx}\rangle\Big)^2}
{\langle\bsy u,\bsy u\rangle(\xi+c\,\langle\bsy u,\bsy u\rangle)\,\xi^2}\right)\bsy u_x+3 (\bsy u_x,\bsy u_{xx})\bsy u,\n 
\end{align}
\begin{align}
& \xi=\langle\bsy u,\bsy u\rangle\langle\bsy u_x,\bsy u_x\rangle -\langle\bsy u,\bsy u_x\rangle^2, \n \\[2mm]
%
\label{A19} \tag{A.19}  
&\bsy u_t= \bsy u_{xxx}+\frac{3}{2}\left(\frac{\langle\bsy u,\bsy u\rangle\langle\bsy u_x,\bsy u_{xx}\rangle -\langle\bsy u,\bsy u_x\rangle\langle\bsy u,\bsy u_{xx}\rangle }{\mu\,\big(\mu
+\langle\bsy u,\bsy u\rangle\big)}-2\frac{\langle\bsy u,\bsy u_x\rangle}{\mu}\right)\big(\bsy u_{xx}+\bsy u_x^2\bsy u\big) \\[2mm]
&\qquad+\frac{3/2}{\langle\bsy u,\bsy u\rangle\big(\mu+\langle\bsy u,\bsy u\rangle\big)} \left[\mu^{-2}\Big(\langle\bsy u,\bsy u\rangle\langle\bsy u_x,\bsy u_{xx}\rangle 
-\langle\bsy u,\bsy u_x\rangle\langle\bsy u,\bsy u_{xx}\rangle \Big)^2\right. \n\\[2mm]
&\hspace{13em}+\langle\bsy u,\bsy u\rangle\langle\bsy u_{xx},\bsy u_{xx}\rangle -\langle\bsy u,\bsy u_{xx}\rangle ^2 \n \\
 &\qquad- 2\,\mu^{-2}\langle\bsy u,\bsy u_x\rangle\Big(\langle\bsy u,\bsy u\rangle\langle\bsy u_x,\bsy u_{xx}\rangle 
-\langle\bsy u,\bsy u_x\rangle\langle\bsy u,\bsy u_{xx}\rangle \Big)\Big(\mu+2\langle\bsy u,\bsy u\rangle\Big)\bigg]\bsy u_{x} \n \\[2mm]
&\qquad+\Big(6\,\mu^{-2}\langle\bsy u,\bsy u_x\rangle^2-3\,\langle\bsy u,\bsy u\rangle^{-1}\langle\bsy u,\bsy u_{xx}\rangle \Big) \bsy u_{x}+3\,(\bsy u_x,\bsy u_{xx})\bsy u,\n \\[1mm]
&\qquad\qquad\mu^2=\langle\bsy u,\bsy u_x\rangle^2+\langle\bsy u,\bsy u\rangle^2-\langle\bsy u,\bsy u\rangle\langle\bsy u_x,\bsy u_x\rangle. \n
\end{align}
Here $a,b$ are arbitrary constants.

\rem 
For equations (\ref{A1}), (\ref{A2}) and  (\ref{A4})  rescaling of the form  $x\to\lambda  x, t\to\lambda^3t$ bring the parameter $a$ to $1$. Moreover, the limit $a\to\infty$ is possible in (\ref{A1}).
The  parameters in equations (\ref{A10}), (\ref{A13}), (\ref{A16}), (\ref{A17}) and (\ref{A18}) can be normalized by   transformations of the form
$\langle\ ,  \rangle\to\lambda \langle\ , \rangle$.  So, without loss of generality we may assume that each parameter in the list (\ref{A1}) -- (\ref{A19})  equals 1 or 0.


\end{document}